# Bike Share's Impact on COVID-19 Transmission and Bike Share's Responses to COVID-19: A case study of Washington DC

Beigi, Pedram[1,3], Haque, Mohaiminul[1], Rajabi, Mohammad Sadra[1], Hamdar, Samer [2]
[1] PhD Student, Department of Civil and Environmental Engineering, The George Washington University, Washington DC, United States
[2] Associate Professor, Department of Civil and Environmental Engineering, The George Washington University, Washington DC, United States
[3] beigi@gwu.edu

**Abstract:** Due to the wide-ranging travel restrictions and lockdowns applied to limit the diffusion of the SARS-CoV2 virus, the coronavirus disease of 2019 (COVID-19) pandemic has had an immediate and significant effect on human mobility at the global, national, and local levels. At the local level, bike-sharing played a significant role in urban transport during the pandemic since riders could travel outdoors with reduced infection risk. However, based on different data resources, this non-motorized mode of transportation was still negatively affected by the pandemic (i.e., relative reduction in ridership). This study has two objectives: 1) to investigate the impact of the COVID-19 pandemic on the numbers and duration of trips conducted through a bike-sharing system – the Capital Bikeshare in Washington, DC, USA; and 2) to explore whether land use and household income in the nation's capital influence the spatial variation of ridership during the pandemic. Towards realizing these objectives, this research looks at the relationship between bike sharing and COVID-19 transmission as a two-directional relationship rather than a one-directional causal relationship. Accordingly, this study models i) the impact of COVID-19 infection numbers and rates on the use of the Capital Bikeshare system and ii) the risk of COVID-19 transmission among individual bike-sharing users. In other words, we examine i) the cyclist's behavior as a function of the COVID-19 transmission evolution in an urban environment and ii) the possible relationship between the bike share usage and the COVID-19 transmission through adopting a probabilistic contagion model. The findings show the risk of using a bike-sharing system during the pandemic and whether bike sharing remains a healthier alternative mode of transportation in terms of infection risk.

## 5. CONCLUSION

In this paper, the impact of Covid-19 on the D.C. bike-share system as well as the bike-share usage impact on disease transmission is explored D.C.'s Capital Bikeshare program lost around 10,000 daily users at the beginning of the Pandemic in March of 2020. Later on, the ridership numbers increased, but the Capital Bikeshare services are still below their daily averages from earlier years by an estimated 9,000 trips per day. The data reveals continued growth in the average trip duration from 13 minutes to 19 minutes. This could mean that people prefer short trips on foot instead of cycling as before the pandemic. Moreover, the stations with the lowest ratio of ridership after to before the pandemic were focused around downtown D.C. The office employment density is a significant factor in bike ridership. While it is reasonable to assume that the upper class of the society use personal vehicles during the COVID-19 and that bicycle riding declines in these areas while the opposite is true in low-income communities, regression analysis shows that there is no significant relationship between wealth, population density and bicycle ridership before and after the pandemic. However, it has been shown that land use plays an essential role in bike-sharing ridership, which it reduced more significantly in commercial and office areas in downtown D.C.

With such usage, to analyze the impact of bike-sharing on the dynamics of the Covid-19 reported case, the transmission process is classified into two types: i) Human-to-human transmission and ii) Human-to-surface-to-human transmission. A probabilistic contagion model is used to represent the first form of transmission. Transmission from one person to another would be infrequent. It's exceedingly improbable that two strangers at a station will come into close proximity at the same time. On the other hand, the second type is more likely to happen, and it will be anticipated using the average number of trips per bike throughout the study interval and the corresponding duration during which the virus remains stable on a given surface.

The numerical analysis indicated that one infected person could infect a maximum of 5 persons, and three infected persons can infect a maximum of 28 persons during a 3-day interval (ignoring the possible infections beyond the bikers' community). The peak value of the 7-day moving average of infected persons is 194.

The findings show that bike-sharing has a minor impact on the Covid-19 infection rate, and that the government and decision-makers should consider it as a safe mode of transportation that should be maintained because it can encourage people to use it instead of the subway or bus. In comparison to other modes of public transit, bike-share remains a relatively healthier option.

The limitation is that there does not appear to be any real-world assessment of how bike-share users interact at stations. Also, for future studies, given the more specific data sets (i.e., the trip history dataset with a user identification number, socio-economic characteristics in a given location, datasets provided at this stage with routing information), the interaction of users at a station can be modeled much more realistic. Moreover, with the travel paths data complemented by pedestrian density information along some routes, we can predict the total impact of bike-share services at the origins, destinations, and in between the origin-destination (O.D.) pairs.